\documentclass[twocolumn]{aastex631}

\usepackage{amsmath}

\shorttitle{MASTER J0302: a DN at a massive ONe WD system ?}
\shortauthors{Kimura et al.}
\graphicspath{{./}{figures/}}

\begin{document}

\title{MASTER OT J030227.28+191754.5: a dwarf nova at a massive oxygen-neon white-dwarf system ?}

\author{Mariko Kimura}
\affiliation{Cluster for Pioneering Research, \\
Institute of Physical and Chemical Research (RIKEN), \\
2-1 Hirosawa, Wako, Saitama 351-0198, Japan}

\author{Kazumi Kashiyama}
\affiliation{Astronomical Institute, Tohoku University, \\
Sendai 980-8578, Japan}
\affiliation{Kavli Institute for the Physics and Mathematics of the Universe, \\
The University of Tokyo, Kashiwa 277-8583, Japan}

\author{Toshikazu Shigeyama}
\affiliation{Research Center for the Early Universe, \\
Graduate School of Science, University of Tokyo, \\
Bunkyo-ku, Tokyo 113-0033, Japan}

\author{Yusuke Tampo}
\affiliation{Department of Astronomy, Graduate School of Science, \\
Kyoto University, Kitashirakawa-Oiwake-cho, Sakyo-ku, \\
Kyoto 606-8502, Japan}

\author{Shinya Yamada}
\affiliation{Department of Physics, Rikkyo University, \\
3-34-1 Nishi-Ikebukuro, Toshima-ku, Tokyo 171-8501, Japan}

\author{Teruaki Enoto}
\affiliation{Cluster for Pioneering Research, \\
Institute of Physical and Chemical Research (RIKEN), \\
2-1 Hirosawa, Wako, Saitama 351-0198, Japan}
\affiliation{Department of Physics, Graduate School of Science, \\
Kyoto University, Kitashirakawa-Oiwake-cho, Sakyo-ku, \\
Kyoto 606-8502, Japan}



\begin{abstract}

We present timing and spectral analysis results of the {\it NICER} and {\it NuSTAR} observations of the dwarf nova MASTER OT J030227.28$+$191754.5 during the 2021--2022 outburst. 
The soft X-ray component was found to be dominated by blackbody radiation with a temperature of $\sim$30~eV and also showed prominent oxygen and neon emission lines. 
The blackbody luminosity exceeded 10$^{34}$~ergs~s$^{-1}$, which is consistent with theoretical predictions, and then decreased more than an order of magnitude in 3.5 days. 
The inferred abundances of oxygen and neon in the optically-thin coronal region surrounding the central white dwarf (WD) are several times higher than the respective solar values. 
\textcolor{black}{Although inconclusive, the abundance enrichment may originate from the WD, indicating that it may be mainly composed of oxygen and neon}. 
Assuming that the blackbody radiation comes from the belt-shaped boundary layer between the WD and the accretion disk, we estimated the WD radius to be $(2.9\pm1.1)\times10^{8}$~cm, which corresponds to the WD mass range of 1.15--1.34~$M_{\odot}$. 
If the accretion continues for another $\sim$Gyr, the WD may experience an accretion-induced collapse into a neutron star and form a so-called black-widow pulsar system.

\end{abstract}



\section{Introduction} \label{sec:intro}

MASTER OT J030227.28$+$191754.5 $=$ AT2021afpi 
$=$ PNV J03022732$+$1917552 (hereafter referred to 
as\ MASTER J0302) is the cataclysmic variable (CV) that 
entered an outburst at the end of 2021 November.
It was discovered on November 26.82568 UT \citep{zhi21afpi} 
as a possible optical counterpart of the neutrino event 
IceCube-211225A, which occurred at 06:22:21.56 UT 
on November 25 \citep{lag21ic211125A}.
Subsequently, an X-ray observation of the object 
was made on November 28 and a significant brightening 
was detected \citep{pal21afpi}. 

According to the Central Bureau for Astronomical Telegrams 
(CBAT), the object had been fainter than 16 mag 
on November 23.7421 UT before the outburst.
Its optical magnitude at the outburst maximum was 
11.4 mag in the $V$ band (Tampo et al.~in prep.), 
whereas the quiescent optical magnitude is 21.94(7) 
in the $g'$ band \citep{ahu20gaia}. 
The large-amplitude outburst and detection of 
the initial optical spectrum dominated by a blue 
continuum with strong single-peaked emission lines 
are initially believed to indicate that this object 
would be a classical He/N nova 
\citep{ste21afpi,boe22neutrino}.
However, the object was later identified as 
a WZ Sge-type system, a subclass of dwarf novae (DNe) 
in further follow-up optical spectroscopy and photometry 
\citep{iso21afpi}. 

The DN is a subgroup of CVs, which is a close binary 
system composed of a ``primary'' of white dwarf (WD) 
and a ``secondary'' of low-mass star.
The WD in a CV is surrounded by an accretion disk 
formed by the transferred mass from the secondary star 
\citep{war95book}.
The accretion rate from the disk to the WD is low 
in the quiescent state.
Due to thermal instability in the accretion disk, 
the accretion rate becomes high, and DNe show sudden 
brightening in the disk, which is called an outburst 
\citep{osa96review,osa05DImodel}.
WZ Sge-type DNe are characterized by large-amplitude 
(typically \textcolor{black}{6--9} mag) outbursts, rebrightening 
after a main outburst, and short-term optical 
modulations, such as early and ordinary superhumps, 
during outbursts \citep[see][for a review]{kat15wzsge}.
Their binary mass ratio ($q \equiv M_2/M_1$, where 
$M_1$ and $M_2$ are the masses of the primary and 
secondary stars, respectively) is extremely low, 
$\leq 0.08$.
\citet{iso21afpi} reported the emergence of early 
superhumps two days after the outburst maximum 
(see the left panel of Figure \ref{earlySH-spectrum}).
Early superhumps appear when the disk radius exceeds 
the 2:1 resonance radius, which is $\sim$1.5-times 
larger than the disk radius in quiescence \citep{osa02wzsgehump}. 
At the onset of the thermal instability, the disk expands 
\textcolor{black}{because of the angular momentum transfer 
from the inner part of the disk to its outer part}.
The emergence of early superhumps is evidence of 
a sudden increase in the mass accretion rate.
At that time, double-peaked emission lines from 
the disk were also observed in the optical spectrum 
(see the right panel of Figure \ref{earlySH-spectrum}). 
These two observations identified MASTER J0302 
as a DN.

The identification as a DN is not consistent with 
the neutrino detection (IceCube-211225A) in the current DN 
and neutrino-emission models, though potential neutrino 
emission from a classical nova is proposed 
\citep{gue23novaeneutrino,bed22novaeneutrino}.
In fact, MASTER J0302 brightened in the optical band 
8.5 hours before the IceCube event \citep{sar21afpi}, 
and thus they must not be associated. 
Two bright AGNs were listed later as the possible 
counterpart of IceCube-211225A \citep{kad21ic211125a}. 
In addition, $\gamma$-ray observations reported 
no significant event \citep{qui21afpi,aya21afpi}; 
the fact is consistent with the identification of 
MASTER J0302 as a DN and not as a neutrino event, 
which is often associated with gamma-ray events.
As such, IceCube-211225A is currently believed 
not to originate from MASTER J0302.

Although previous studies of DNe have elucidated 
their optical properties, many unsolved problems remain 
in their X-ray properties. 
What we know about the X-ray properties and what 
we don't are briefly summarized below. 
In the current standard model of the DN, its X-ray 
emitting plasma is located in the vicinity of the WD 
and is powered by accretion.
When the system enters an outburst, the accretion rate 
becomes higher than usual so that the total energy 
released by accretion becomes larger. 
During the outburst, the optically-thin X-ray plasma 
in quiescence becomes optically thick  
\citep{pat85CVXrayemission1,nar93BL,pat85CVXrayemission2}.
As a result, the X-ray luminosity in general decreases 
during the outburst and in contrast, the extreme-ultra-violet 
(EUV) luminosity increases.
However, several DNe have been reported to show 
an increase in X-ray flux during outbursts, and 
WZ Sge-type DNe belong to these outliers 
\citep{whe05wzsge,byk09gwlibSwift,neu18j1222}. 
Although a naive expectation from the standard model 
is that the observed high X-ray flux during outbursts 
may be attributed to soft X-rays radiated from 
the optically-thick boundary layer, which would be 
a blackbody emission with a temperature of several 
tens of eV, such blackbody emissions were observed 
only in a handful of objects during outbursts 
\citep{mau04sscyg,byk09gwlibSwift}. 
In general, the observed soft X-ray luminosity 
during outbursts is more than 100 times weaker 
than the prediction. 
This ``missing boundary-layer problem'' proposed by 
\citet{fer82missingBL} four decades ago remains 
unsolved so far. 

Another unsolved problem is the ultimate fate 
in the evolution of CVs. 
It has long been proposed that some CVs will eventually 
evolve into (neutron-star) low-mass X-ray binaries 
\citep{kul88lmxb}; 
Observationally, WZ Sge-type DNe would be the best 
targets to constrain the hypothesis, given that 
WZ Sge-type DNe are believed to be the most evolved 
CV systems. 
In particular, their binary parameters are keys.
WZ Sge-type DNe are too faint to observe during 
quiescence due to their low mass-transfer and 
mass-accretion rates. 
Their outbursts provide good opportunities for 
determining their binary parameters observationally. 
X-ray observations of WZ Sge-type DNe during their
outbursts are no exceptions and they have two specific 
advantages: to tackle the missing boundary-layer problem 
through search for the soft X-ray emission and 
to constrain the evolutionary path of CVs possibly 
by means of determination of the WD mass and/or abundances
of elements in the high-temperature plasma.

MASTER J0302 showed the largest outburst among 
WZ Sge-type DNe and became brighter in X-rays 
during outburst, as other WZ Sge-type DNe do.
Then, its 2021--2022 outburst provides us with 
a good opportunity to understand the nature of 
the X-ray plasma in DNe and the WD properties 
of this source. 
In this paper, we present the analysis results of 
\textit{NICER} and \textit{NuSTAR} data of MASTER J0302 
taken during the outburst. 
To detect supersoft X-rays from the boundary layer is 
unfeasible for conventional X-ray detectors 
for monitoring use since most of which are sensitive 
only in hard X-rays above $\sim$1~keV, and since 
large X-ray satellites seldom respond to alerts 
for outburst events in a timely manner. 
On the other hand, {\it NICER} is suitable for 
detecting the transient supersoft X-ray emission 
because it covers an energy range down to 0.2~keV 
and is well capable of flexible observations of 
transients. 
Hard X-ray observations with {\it NuSTAR} are also 
important to investigate the emission from 
the optically-thin coronal flow that coexists with 
the boundary layer even during outbursts 
\citep{wad18gkper,kim21sscyg}.
This paper is organized as follows. 
Section 2 describes the X-ray observations.
Section 3 shows our analysis results of light curves and 
spectra.
We give our interpretation of the object and its X-ray 
radiation mechanism in section 4.

Throughout this paper, we assume the binary mass 
ratio $q = 0.063$ and the inclination angle 
$i = 60\pm10\,\mathrm{deg}$ for MASTER J0302 
on the basis of the analysis of the optical 
data~(Tampo et al.~in prep).
We also assume the distance to the source 
$d = 720\,\mathrm{pc}$~\citep{kat22standardcandle}, 
which was determined from the disk component at 
the superhump as a standard candle and has 
an uncertainty of $\sim$23\%.

\section{Observations and analyses} \label{sec:obs}

\subsection{NICER}

{\it NICER} monitored MASTER J0302 during 
its 2021--2022 outburst, starting 1.6~days after 
the epoch of the optical maximum. 
The  observation IDs (ObsIDs) of the monitoring 
are given in Table \ref{nicer-log} in the appendix section. 
In this work, we use HEAsoft version 6.30.1 for 
data reduction and analyses. The data were reprocessed 
with the pipeline tool \texttt{nicerl2}, which used  
the {\it NICER} Calibration Database (CALDB) version 
later than 2021 July 20, for producing light curves 
and time-averaged spectra. 
The light curves were generated by \texttt{lcurve}.
The source and background spectra were extracted with 
\texttt{nibackgen3C50} version 7. 
{\it NICER} is composed of 56 modules of silicon drift 
detectors (SDDs), 52 of which are operating 
on orbit, including two noisy modules, IDs 14 and 34. 
In our data reduction with \texttt{nicerl2}, 
we filtered out the data of the noisy two modules. 
For spectral analyses, we obtained the response matrix 
file and the ancillary response file for a specific set 
of 50 detectors to match the default settings of 
the background model.\footnote{The method is described at $<$https://heasarc.gsfc.nasa.gov/docs/nicer/analysis\_threads/arf-rmf/$>$, and we use the additional data version xti20200722.}
All of the observation times were converted to 
barycentric Julian date (BJD) by \texttt{barycorr}.

\subsection{NuSTAR}

The {\it NuSTAR} Target of Opportunity (ToO) observations 
were carried out  only on 2021 December 2. 
The observation log is given in Table \ref{nustar-log}. 
The observed area was partially overlapped with that of 
the {\it NICER} observation of ObsID 4202450104 
on the same day. 
The data were reprocessed  with \texttt{nupipeline} and 
the {\it NuSTAR} CALDB as of 2021 October 20.  
The light curves, time-averaged spectra, and response and 
ancillary response files  were obtained  with \texttt{nuproducts}.  
The background region was circular with a radius of 62'' 
at a blank sky area.  
We determined the circular source region centered at 
the target position with a 25'' radius.

\section{Results}

\subsection{Light curves and variability} \label{sec:overall}

\begin{figure*}[htb]
\epsscale{0.7}
\plotone{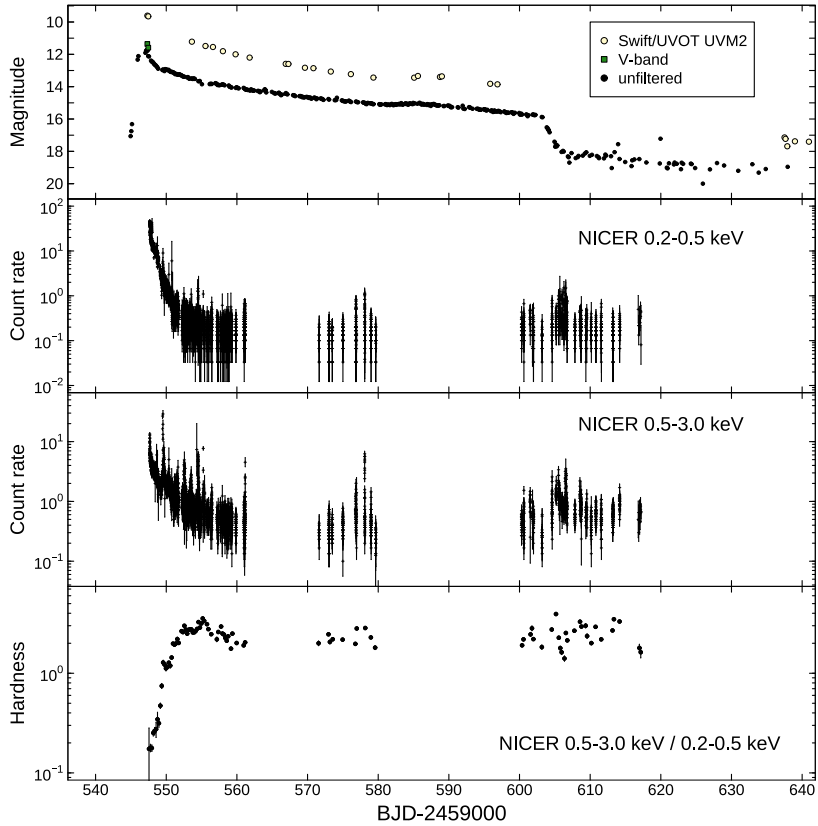}
\caption{
UV, optical, and X-ray light curves of the 2021--2022 outburst of MASTER J0302 for the entire observation period.  
Top panel shows the optical (green squares) $V$-band, (black filled circles) unfiltered, and (white open circles) {\it Swift} UM2-band data. 
The Variable Star Network (VSNET) team provides the optical data. 
The central wavelength of the {\it Swift} UM2-band is 2246 \AA. 
The second and third panels show {\it NICER} X-ray light curves binned into 30-s bins in the 0.2--0.5~keV and 0.5--3.0~keV bands, respectively. 
Bottom panel shows the hardness ratio, the 0.5--3.0 keV count rate divided by the 0.2--0.5 keV count rate.
}
\label{overall}
\end{figure*}

The ultraviolet (UV), optical, and X-ray light curves 
are displayed in Figure \ref{overall}. 
The optical light curves clearly show that the outburst 
lasted for $\sim$2 months. 
\textcolor{black}{The outburst amplitude was $\sim$10 mag, 
which was unusually large in comparison with those 
of normal WZ Sge-type DNe, though there are several 
WZ Sge-type systems that exhibited outbursts with 
$>$9-mag amplitudes \citep{kat15wzsge,tam20j2104}.
Using the same method described in section 2.1 of 
\citet{dub18DItest}, we derived the mass accretion 
rate ($\dot{M}_{\rm acc}$) of the disk onto the WD  
to be $\dot{M}_{\rm acc} \sim 10^{19}$~g~s$^{-1}$ 
($= \sim 2 \times 10^{-7}$~$M_{\odot}$~yr$^{-1}$) 
to reproduce $V$ = 12 mag, the observed brightest 
optical magnitude of the source (top panel of 
Figure \ref{overall}), assuming the standard disk 
with a uniform accretion rate over the entire disk, 
$R_{\rm disk}$ = 3$\times$10$^{10}$~cm, 
$M_1$ = 1.2~M$_{\odot}$, and $R_{\rm in}$ = 
5$\times$10$^{8}$~cm, where $R_{\rm disk}$ 
and $R_{\rm in}$ are the radii of the outer and 
inner disk edges, respectively.
Here, we assume that $R_{\rm disk}$ reached the 2:1 
resonance radius \citep{osa02wzsgehump}.
The estimated accretion rate is higher than predicted 
for normal WZ Sge-type outbursts \citep{osa95DNproc}, which 
would be attributed to the high-amplitude outburst.}

The decline rate of optical light curves is consistent 
with those of other WZ Sge-type DNe 
\citep{kat15wzsge}.
The decline of the UV light curve was similar to that 
of the optical light curve.
The {\it NICER} observation started on 2021 November 28 
(BJD 2459547), 1.6~d after the optical outburst maximum.
The observed X-ray flux was highest 
at 3.8$\times$10$^{-11}$~ergs~s$^{-1}$cm$^{-2}$ 
in the 0.2--3.0 keV when the observations started and 
then it rapidly declined.
Considering the uncertainty of the distance to MASTER J0302, 
the optical and X-ray luminosities are less than 
4.6$\times$10$^{35}$~ergs~s$^{-1}$ and 
9.3$\times$10$^{34}$~ergs~s$^{-1}$, respectively, 
which are about three orders of magnitude lower than 
the Eddington luminosity. 
This suggests that thermonuclear runaway on the WD surface 
is unlikely to have occurred at this object.
The X-ray flux declined faster than the optical one and 
decreased by 70\% and 30\% per day in the 0.2--0.5~keV and 
0.5--3.0~keV bands, respectively, till BJD 2459551. 
The hardness ratio of the fluxes between these two energy 
ranges, i.e., (0.5--3.0~keV)/(0.2--0.5~keV), 
rapidly increased as the X-ray fluxes decreased 
before it became stable eventually.

The X-ray light curve appears to have some spiky, 
potentially flaring, events (second and third panels 
of Figure \ref{overall}). 
However, they are likely to originate from background 
fluctuations (see also Figure \ref{variability}), 
as we found that each of the spiky events was synchronized 
with a sharp increase in the rate of the so-called 
``overshoot'' events in {\it NICER}. 
Overshoot events are background events caused by 
a number of charges deposited along their traveling 
paths in the {\it NICER} SDD when high-energy charged 
particles, such as cosmic rays, enter the SDD\footnote{For more details about the {\it NICER} overshoot event, see \url{https://heasarc.gsfc.nasa.gov/docs/nicer/analysis_threads/overshoot-intro/}}.

\subsection{Time-dependent X-ray spectra} \label{sec:spectra}

\begin{figure*}[htb]
\begin{minipage}{0.49\hsize}
\plotone{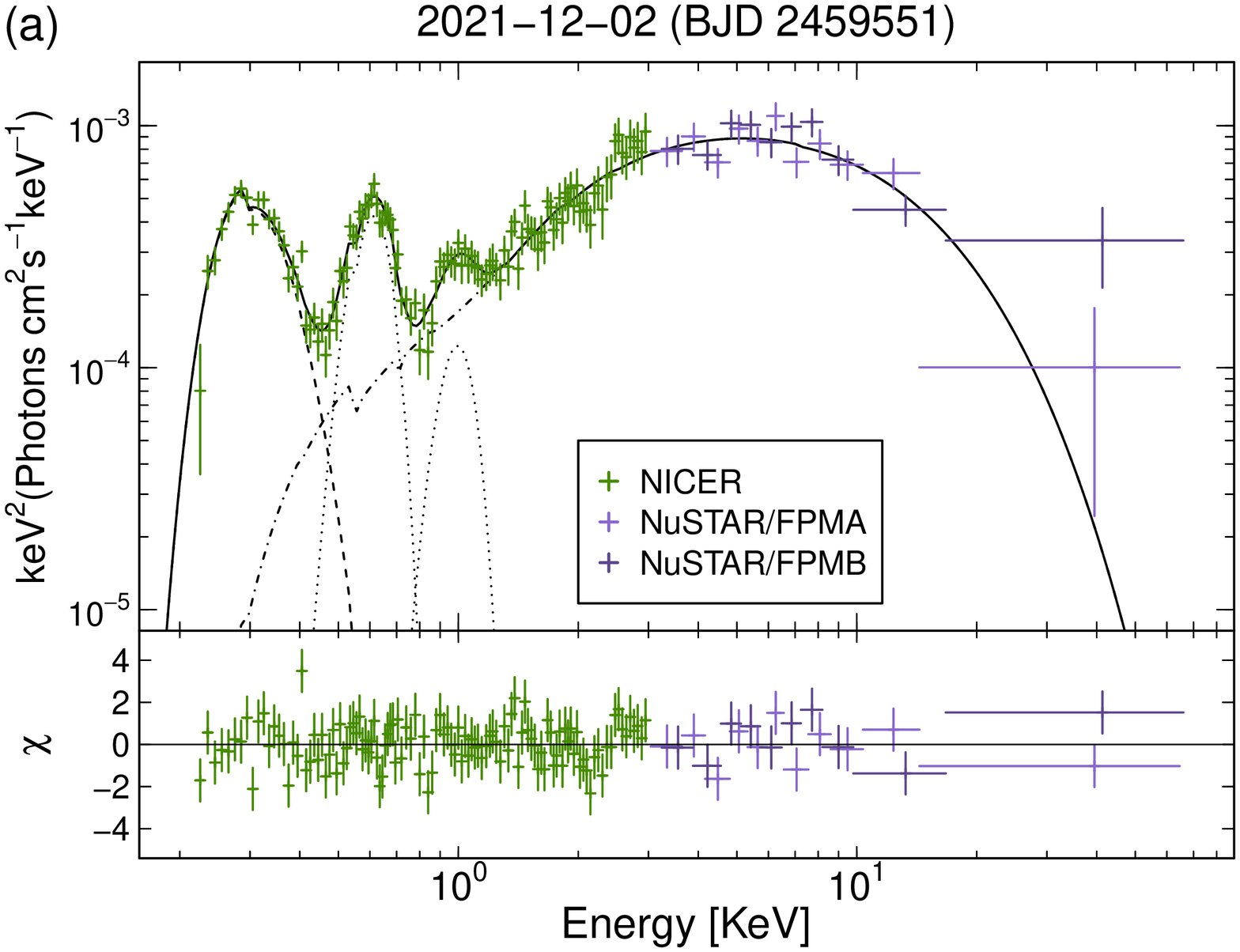}
\end{minipage}
\begin{minipage}{0.49\hsize}
\plotone{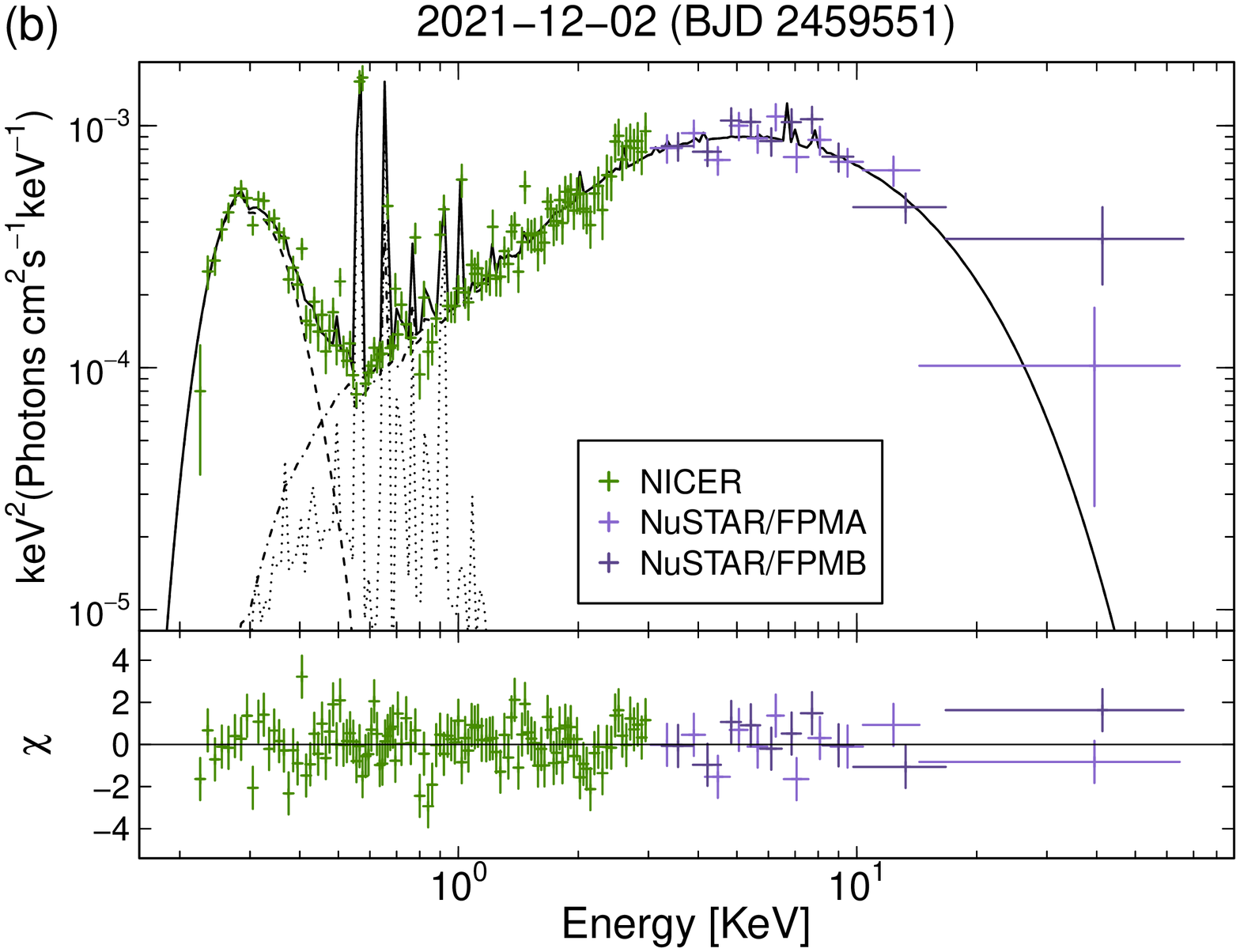}
\end{minipage}
\caption{
Broad-band X-ray SEDs of MASTER J0302 on BJD 2459551 (2021 December 2) during the outburst, overlaid with the best-fit spectral models of (a) \texttt{Tbabs*pcfabs*(bbody+gaussian+gaussian+bremsstrahlung)} and (b) \texttt{Tbabs*pcfabs*(bbody+vapec+vapec)}. Green crosses represent the {\it NICER} data. 
Purple and dark purple crosses represent the {\it NuSTAR} FPMA and FPMB data, respectively. 
In panel (a), dashed line and dot-dashed line represent the best-fit model components  of blackbody and bremsstrahlung, respectively, and the two dotted lines do those of  the oxygen and neon lines.
Similarly, in panel (b), dashed line, dotted line, and dot-dashed line represent   those of blackbody, low-temperature and high-temperature collisionally-ionized plasma emissions, respectively. Solid line in each panel shows the total best-fit model emission. The normalizations of the {\it NuSTAR} SED were 0.81 and 0.90 of the {\it NICER} SED for the data in panels (a) and (b), respectively. 
}
\label{spec211202}
\end{figure*}

\begin{table*}[htb]
    \caption{
    Best-fit parameters for models of  (a) \texttt{Tbabs*pcfabs*(bbody+gaussian+gaussian+bremsstrahlung)}  and  (b) \texttt{Tbabs*pcfabs*(bbody+vapec+vapec)} in the simultaneous spectral model-fitting of the simultaneous {\it NICER} and {\it NuSTAR} observation data of MASTER J0302 on BJD 2459551 (2021\ December 2).  
    The errors represent 90\% confidence ranges.  
    }
    \vspace{2mm}
    \label{parameter-211202}
    \centering
\begin{tabular}{cccc}
\hline
Model & Parameter & (a) & (b) \\
\hline
pcfabs & $N_{\rm H}$$^{*}$ 
      & 2.7$_{-1.7}^{+1.0}$ & 1.7$_{-0.6}^{+1.1}$ \\
      & $f$$^{\dagger}$
      & 0.62$_{-0.09}^{+0.05}$ & 0.61$_{-0.07}^{+0.06}$ \\
\hline
bbody & $T_{\rm BB}$$^{\ddagger}$ 
      & 30$_{-0.5}^{+1.0}$ & 30$\pm$1.1 \\
      & $L_{\rm BB}$$^{\S}$ 
      & 1.5$\pm$0.4 & 1.3$_{-0.3}^{+0.5}$ \\
\hline
gaussian 1 & $E_1$$^{\P}$
		& 0.59$\pm$0.006 & -- \\
		& $\sigma_1$$^{|}$ 
        & 6.0$_{-0.7}^{+1.0}$$\times$10$^{-2}$ & -- \\
        & N$_1$$^{**}$ 
        & 7.4$_{-1.5}^{+1.3}$ & -- \\
\hline
gaussian 2 & $E_2$$^{\P}$ 
		& 0.98$\pm$0.03 & -- \\
		& $\sigma_2$$^{|}$
        & 5.5$_{-2.9}^{+3.0}$$\times$10$^{-2}$ & -- \\
        & N$_2$$^{**}$ 
        & 5.0$_{-2.2}^{+7.7}$$\times$10$^{-1}$ & -- \\
\hline
vapec 1 & $kT_{1}$$^{\dagger\dagger}$
		& -- & 0.18$_{-0.01}^{+0.01}$ \\
		& $Z_{\rm O}$$^{\ddagger\ddagger}$ 
        & -- & 5.8$_{-3.1}^{+5.9}$ \\
        & $Z_{\rm Ne}$$^{\S\S}$ 
        & -- & 12$_{-6.8}^{+12}$ \\
        & $Z_{\rm Fe}$$^{\P\P}$ 
        & -- & $\leq$0.34 \\
        & N$_1$$^{||}$
        & -- & 1.3$_{-0.6}^{+1.4}$$\times$10$^{-4}$ \\
\hline
bremsstrahlung & $T_{\rm bremss}$$^{\dagger\dagger}$ 
      & 6.7$_{-0.9}^{+1.0}$ & -- \\
      & $N_3$$^{***}$ 
      & 8.6$_{-0.9}^{+1.0}$$\times$10$^{-4}$ & -- \\
\hline
vapec 2 & $kT_{2}$$^{\dagger\dagger}$ 
		& -- & 6.3$_{-0.8}^{+0.9}$ \\
        & N$_2$$^{||}$ 
        & -- & 1.5$_{-0.4}^{+0.6}$$\times$10$^{-3}$ \\
\hline
$\chi^2$/dof & & 1.08 & 1.14 \\
\hline
\multicolumn{4}{l}{\parbox{250pt}{$^{*}$Equivalent hydrogen column in 10$^{22}$ atoms cm$^{-2}$.}}\\
\multicolumn{4}{l}{\parbox{250pt}{$^{\dagger}$Dimensionless covering fraction ($0 < f \leq 1$).}}\\
\multicolumn{4}{l}{$^{\ddagger}$Temperature in eV.}\\
\multicolumn{4}{l}{\parbox{250pt}{$^{\P}$Blackbody luminosity in units of 10$^{34}$~ergs~s$^{-1}$.}}\\
\multicolumn{4}{l}{$^{\S}$Line energy in keV.}\\
\multicolumn{4}{l}{$^{|}$Line width in keV.}\\
\multicolumn{4}{l}{\parbox{250pt}{$^{**}$Total photons/cm$^2$/s in the line of sight in units of 10$^{-4}$.}}\\
\multicolumn{4}{l}{$^{\dagger\dagger}$Plasma temperature in keV.}\\
\multicolumn{4}{l}{\parbox{250pt}{$^{\ddagger\ddagger}$Oxygen abundance with respect to the solar one.}}\\
\multicolumn{4}{l}{\parbox{250pt}{$^{\P\P}$Neon abundance with respect to the solar one.}}\\
\multicolumn{4}{l}{\parbox{250pt}{$^{\S\S}$Iron abundance with respect to the solar one.}}\\
\multicolumn{4}{l}{\parbox{250pt}{$^{||}$$\frac{10^{-14}}{4\pi (D_{\rm A} (1 + z))^2}\int n_e n_I dV$, where $D_{\rm A}$ is the angular distance to the source (cm), $dV$ is the volume element (cm$^3$), and $n_e$ and $n_H$ are the electron and hydrogen densities (cm$^{-3}$), respectively.}}\\
\multicolumn{4}{l}{\parbox{250pt}{$^{***}$$\frac{3.02\times10^{-15}}{4\pi D^2} \int n_e n_I dV$, where $D$ is the distance to the source (cm).}}\\
\end{tabular}
\end{table*}

The broadband X-ray spectrum was taken on BJD 2459551 
by \textit{NICER} and \textit{NuSTAR}. 
First, we fitted the spectra with the model 
\texttt{Tbabs*pcfabs*(bbody+gaussian+gaussian+bremsstrahlung)} 
in the \texttt{XSPEC} software \citep{XSPEC}, 
where \texttt{Tbabs}, \texttt{pcfabs}, \texttt{bbody}, 
\texttt{gaussian}, and \texttt{bremsstrahlung} 
denote X-ray absorption by the interstellar medium, 
partial-covering fraction for the absorption of X-rays, 
blackbody radiation, a single Gaussian model typically 
representing an emission line, and optically-thin 
bremsstrahlung radiation, respectively. 
The cross-normalization factor between the {\it NICER} 
and {\it NuSTAR} spectral energy distributions (SEDs) 
was corrected in the fitting.
The result is exhibited in the left panel of 
Figure \ref{spec211202}. 
Another form of plot for the same result is given 
in the left panel of Figure \ref{spec211202-ld}, in which the y-axis 
is the normalized count rate.
We  did not use the {\it NICER} spectrum above 3.0 keV 
because the estimated background rate  of 0.1 
counts~s$^{-1}$~keV$^{-1}$ 
at 3.0 keV was comparable to the observed count rate 
at the energy and it was even smaller at a higher energy.
The column density $N_{\rm h}$ was fixed at 
7.3$\times$10$^{20}$~cm$^{-2}$ in the \texttt{Tbabs} model; 
the value had been determined through simultaneous 
model-fitting of several {\it NICER} SEDs with a common 
$N_{\rm H}$ and is consistent with the galactic absorption 
to the direction of MASTER J0302.

We found that the spectrum is composed of three components; 
a blackbody component with a temperature of 
30$\pm$1 eV, two possible emission lines with 
non-negligible widths at around 0.6 and 1.0 keV, 
and bremsstrahlung radiation with a temperature 
of $\sim$7 keV. 
Table \ref{parameter-211202} tabulates 
the best-fit parameters. 
The two broad lines are possibly blended lines 
emitted from helium-like and hydrogen-like ions 
of oxygen and neon, respectively, which {\it NICER} 
cannot resolve. 
Reportedly, the helium-like oxygen emission line 
in GW Lib, a WZ Sge-type DN, which was strong 
during the quiescence state, disappeared during 
its outburst \citep{byk09gwlibSwift,hil07gwlibXMM}. 
The present result provides the first detection of 
strong oxygen and neon lines during an outburst in a DN.
Complex iron emission lines at around 6.5~keV that are 
prominent in several DNe 
\citep{ish09sscygSuzaku,wad17XrayDNe,per03wxhyiChndra,szk02ugemChandraproc} 
were not detected during the outburst in MASTER J0302; 
the lack of detection is consistent with past observations 
of GW Lib and WZ Sge during the early stage of outbursts
\citep{pat98wzsge,hil07gwlibXMM,byk09gwlibSwift}.
We estimated the upper limit of the iron-line flux 
from MASTER J0302 to be 10$^{-13}$~erg~s$^{-1}$~cm$^{-2}$ 
by performing the modeling of simulated spectra.
Here, we generated the simulated {\it NuSTAR} spectra 
in the $\geq$3~keV band with the \texttt{fakeit} command 
in \texttt{XSPEC} on the basis of the model 
\texttt{Tbabs*(bremsstrahlung+gaussian)}, 
where the \texttt{gaussian} model represents the iron 
emission line at 6.5~keV, and investigated the strength 
of the iron line to determine how strong it could 
be observed.

We also performed another spectral modeling with the model 
\texttt{Tbabs*pcfabs*(bbody+vapec+vapec)}.
Here, \texttt{vapec} represents the radiation from 
a collisionally-ionized diffuse gas on the basis of 
the \texttt{AtomDB} database for atomic parameters.
The parameters of oxygen, neon, and iron abundances 
for the two \texttt{vapec} components were assumed 
to be the same for each element and the abundances 
of other elements than these three were fixed 
at the solar values.
The right panel of Figure \ref{spec211202} and 
Table \ref{parameter-211202} show the fitting result 
and best-fit parameters, respectively.
The result implied that the oxygen and neon abundances 
were $\sim$3-times and $\sim$5-times of their solar abundances, 
respectively, whereas the X-ray emitting plasma in most DNe 
has sub-solar abundances for oxygen and neon 
\citep{pan05XrayDNe}. 

\begin{figure}[htb]
\epsscale{1.0}
\plotone{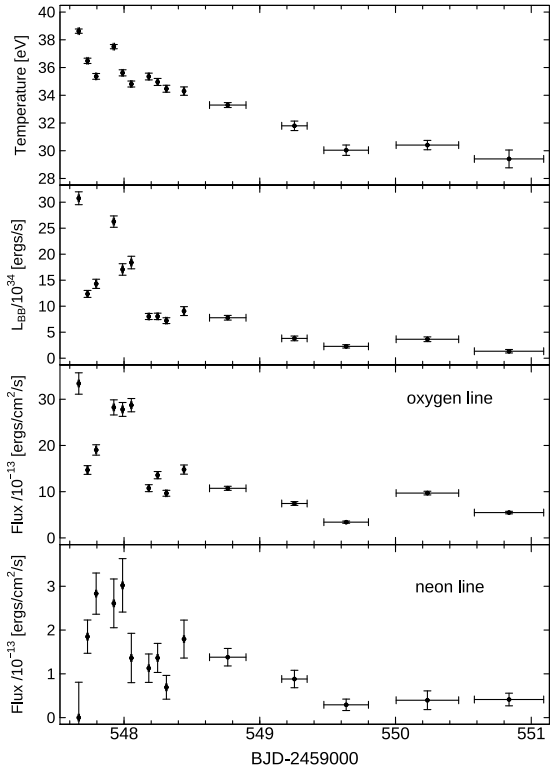}
\caption{
Time evolution of the blackbody radiation, oxygen and neon emission lines from MASTER J0302.
Top and second panels show the time evolutions of the temperature and luminosity, respectively, of the blackbody component. 
The fourth and bottom panels show the time evolutions of the fluxes of the oxygen and neon lines, respectively.
The line fluxes are corrected for absorption.
The errors are in 90\% confidence.
}
\label{spectral-evolution}
\end{figure}

Next, we investigated the time evolutions of the blackbody 
component and oxygen and neon emission lines 
by model-fitting the {\it NICER} data at individual epochs, 
collectively spanning for 3.5 days, with the first of 
the above-used model, 
\texttt{Tbabs*pcfabs*(bbody+gaussian+gaussian+bremsstrahlung)}. 
In the fitting, due to the statistics limitation, we fixed 
the bremsstrahlung temperature, central energies and 
widths of the emission lines, and partial-absorption 
parameters at the best-fit values in Table 
\ref{parameter-211202}. We note that the bremsstrahlung 
temperature of $\sim$7~keV could be hardly constrained 
with this model-fitting of the {\it NICER} data alone 
because its data above 3.0 keV were dominated by 
the background and because no {\it NuSTAR} data were 
available other than those on BJD 2459551.
Figure \ref{spectral-evolution} shows the obtained time 
evolutions of the temperature and luminosity of 
the blackbody component, and the fluxes of 
oxygen and neon emission lines. 

Notice from Figure \ref{spec211202} that the luminosity 
of the blackbody component exceeded $10^{35}$ ergs~s$^{-1}$ 
on BJD 2459547 and then decreased with time.
The blackbody temperature ranged between 30--40 eV 
till BJD 2459551 and then decreased with time. 
The temperature during the outburst was a few times 
higher than those in other DNe. 
The blackbody flux declined faster than the neon and 
oxygen line fluxes. 
There is a clear difference in the observed decline rates 
between the two energy bands and that explains 
the rapid time evolution of the hardness ratio 
at the early stage of the outburst (bottom panel of 
Figure \ref{overall}).
The absorption-corrected oxygen and neon line luminosities 
were as strong as a few times  10$^{32}$ and 10$^{31}$ 
ergs~s$^{-1}$, respectively, around BJD 2459547 
in MASTER J0302, whereas they were known to be less than 
$10^{29}$ ergs~s$^{-1}$ in other DNe 
\citep{per03wxhyiChndra,szk02ugemChandraproc}.

\section{Discussion} \label{sec:discuss}

\subsection{X-ray properties of MASTER J0302 as a dwarf nova}

\begin{figure}[htb]
\epsscale{1.0}
\plotone{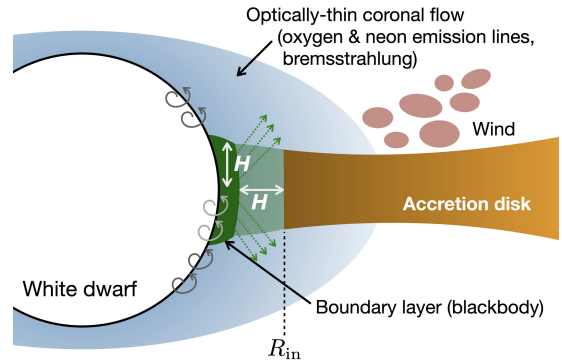}
\caption{
Schematic illustration of the accretion flow in the vicinity of the WD surface in MASTER J0302 in the edge-on view.
Blue, green, and orange regions represent the high-temperature coronal flow observed in hard X-rays, the optically-thick boundary layer (with a scale height of $H$) observed in EUV and soft X-rays, and the low-temperature accretion disk observed at UV and optical wavelengths, respectively. The gas at the WD surface may leak to X-ray emitting plasma. Winds (red regions) may be blowing from the accretion disk and obscure part of soft X-rays emitted from the system.
}
\label{picture}
\end{figure}

The X-ray observational characteristics of MASTER J0302 
are basically consistent with those observed in 
the past from other DNe during outbursts, 
though some unusual features were detected. 
Figure \ref{picture} displays the schematic picture 
of the X-ray emitting plasma in this source.
CVs are broadly categorized into two groups according to 
the strength of the magnetic field of the WD: 
either strongly or weakly magnetized. 
We consider that this object is a non-magnetic CV because 
no significant periodicity originating from the WD 
rotation was found. 
If MASTER J0302 harbored a WD with a strong magnetic field, 
short-term periodic variations would be detected 
due to column accretion onto the spinning WD 
\citep{mau06aeaqr,ken17foaqr}.

In the widely-accepted model, the X-ray emitting region 
in a DN exists in the vicinity of the WD, 
and it becomes optically thick (``boundary layer'') 
during outbursts and emits supersoft X-rays of 
blackbody radiation with a temperature of several 
tens of eV (see Introduction). 
It is usually difficult to detect a soft X-ray 
blackbody component with such a low temperature 
in DNe during outbursts. There have been a limited 
number of successful EUV and soft X-ray detections 
from this kind of objects, for example OY Car and U Gem 
\citep{lon95ugemEUVE,mau00oycarEUVwind,byk09gwlibSwift}.
In this study, we used {\it NICER}, and successfully 
detected a low-temperature blackbody emission mere 
1.6~days after the optical maximum, which had 
a temperature of $\sim$30~eV and a luminosity of 
$\sim$10$^{34}$~erg~s$^{-1}$, in MASTER J0302 
(see Figure \ref{spec211202}). 
The observed luminosity is consistent with that 
from the boundary layer predicted by a theoretical 
model of \citet{fer82missingBL}.

Soft X-rays from the boundary layer in the WD system 
irradiate the inner part of the disk and drive 
outflows (Figure~\ref{picture}). 
The outflow feature is often observed to be 
imprinted in the UV spectroscopy of DNe 
during outbursts \citep{fro01ugem,lon03wzsgeFUSE}.
Our spectral analyses revealed dense gas with $N_{\rm h}$ 
of $\sim$10$^{22}$~cm$^{-2}$ absorbing soft X-rays partially 
covering the X-ray emitting region, which may represent 
outflows and agrees with the general model of DNe 
(Figure~\ref{picture}). 
Also, our results suggest that a multi-temperature 
optically-thin coronal flow coexisted with 
an optically-thick boundary layer, the feature 
of which were observed in other DNe with high 
mass-accretion rates \citep{dob17mvlyr,byk09gwlibSwift}. 
The oxygen and neon emission lines, as we detected 
in MASTER J0302, are also likely to originate from 
the optically-thin corona.

However, the following observed characteristics 
 are unusual for  DNe.
\begin{itemize}
	\item  The oxygen and neon lines were  very strong. 
	The estimated abundances of oxygen and neon are 
	at least a few times higher than their solar values.
	\item The radius of the blackbody emitter was 
	as small as $\sim$4$\times$10$^{7}$~cm, and 
	the blackbody temperature was as high as $\sim$30~eV.
\end{itemize}

\subsection{\textcolor{black}{Possible cause of prominent oxygen and neon emission lines}}

The X-ray emitting plasma in MASTER J0302 was much richer 
in oxygen and neon than that in other DNe 
(right column of Table \ref{parameter-211202}).
There are several possible hypotheses to explain 
the high abundances.
One possibility is (A) the WD core is enriched 
with oxygen and neon, i.e., an ONe WD, and another 
possibility is (B) the secondary star provides 
oxygen and neon-rich gas to the boundary layer 
via the disk accretion.
\textcolor{black}{
Although we cannot completely rule out the latter case (B), 
it seems highly unlikely because the secondary star 
is either a red dwarf or a brown dwarf \citep{hil22novae}.}

\textcolor{black}{
Let us discuss the former case (A) as follows.
In CVs in general, the difference in velocity between 
the accreted matter and the matter on the WD surface 
is expected to induce the Kelvin-Helmholtz instability, 
which stirs matter at the boundary between the WD surface 
and the X-ray emitting plasma. 
However, the WD surface is covered by the accreted 
matter enriched with hydrogen. 
The boundary layer does not obtain oxygen and 
neon-rich gas through the above-mentioned process 
without recent nova eruptions, even if the WD core 
contains gas enriched with these two elements.}

\textcolor{black}{
If a nova eruption occurs, the material of the WD core 
should be dredged up. 
If the WD core is enriched with oxygen and neon, 
the gas enriched with these two elements will be 
ejected and pollute the X-ray emitting plasma, 
the disk, and the secondary star. 
Since oxygen and neon-rich gas is not produced by 
the nuclear burning at the hydrogen-rich layer 
covering the WD, it should originate from 
the WD core \citep{hat16neonnova}. 
A recent nova eruption may form a nova shell, 
though it was not detected in around half of novae 
and its time evolution has a rich variety  
\citep{dow00novaabsmag,tap20novashell}.
The lack of high-resolution images of specific emission 
lines sensitive to nova shells prevents us from 
searching for nova shells.
The possibility of a past nova eruption, therefore, 
still remains.
We estimate the amounts of oxygen and neon gas 
on BJD 2459551 to be $\sim$8$\times$10$^{15}$~g and 
$\sim$3$\times$10$^{15}$~g, respectively, by using 
the best-fit values of the X-ray spectral modeling 
in section 3.2, which are much smaller than the ejected 
mass at one nova eruption. 
In this estimation, we assume that the X-ray emitting 
plasma is spherical with a radius of 10$^{9}$~cm. 
The amount of mass ejected from a massive WD 
at one nova eruption is $\sim$10$^{-7}$ to 10$^{-5}~M_{\odot}$, 
and its $\sim$1--10\% is regarded to be the gas enriched 
with oxygen and neon \citep{pri95nova,yar05nova,hat16neonnova}.
It may not be surprising that the small amount of 
oxygen and neon-rich gas accreted from polluted gas 
in the disk and secondary star. 
To distinguish (A) and (B), we have to estimate 
the metallicity of the gas on the WD surface 
by UV spectroscopy.
To investigate whether a nova shell exists or not, 
we have to perform high-resolution X-ray/UV/optical 
imaging.
These observations and estimations are beyond 
the scope of this paper.}

\subsection{\textcolor{black}{Possible evidence for a massive white dwarf}}

\begin{figure}[htb]
\epsscale{1.0}
\plotone{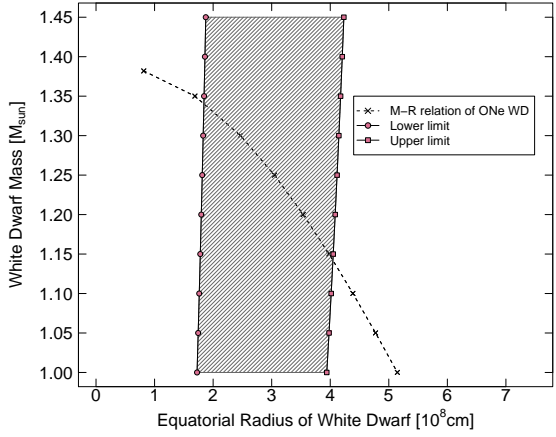}
\caption{
Relation between the mass and radius of the ONe WD. Dashed line with crosses represents the equatorial radius of the ONe WD with no rotation,  as calculated by \citet{kas19magneticwind} (see also Figure 2 and text in  the paper for details). The solid line with circles and that with squares are the lower and upper limits, respectively, of the WD mass as a function of the disk inner-edge radius   for MASTER J0302, which are  calculated according to equation (4) and observed quantities. The shaded region  shows the possible range of the WD mass of MASTER J0302.
}
\label{wd-mass}
\end{figure}

The following discussion is based on the hypothesis 
that the high abundances of oxygen and neon originate 
from the WD gas.
The idea that MASTER J0302 has an ONe WD seems to be 
consistent with the compactness of its WD.
Under the assumption that X-rays from the boundary layer 
are radiated from the surface of a belt-shaped region 
around the equatorial plane of the WD (see the green 
region and arrows in Figure \ref{picture}), 
we estimated the WD radius ($R_1$) from the quantities determined 
by the SED modeling according to the following relation: 
\begin{equation}
4 \pi R_{1} H = \frac{L_{\rm BB}}{\sigma T_{\rm BB}^4}, 
\label{Rin-Rbb}
\end{equation}
where $\sigma$ is the Stefan-Boltzmann constant and 
$H$ is the scale height of the boundary layer, 
which is defined as the vertical extent from the equatorial 
plane to the surface (see Figure \ref{picture}).
Also, $L_{\rm BB}$ and $T_{\rm BB}$ denote 
the luminosity and temperature of the blackbody emitter, 
respectively.
Here, we postulate that the radial extent of 
the boundary layer is $\sim$H \citep{AccretionPower3}. 
Since the inclination angle of this object is less 
than 70 deg, the boundary layer is not eclipsed 
by the disk.
An approximate analytic formula of $H$  
is given by
\begin{equation}
\begin{split}
\left(\frac{H}{R}\right)^2 = 6.96 \times 10^{-4} \left(\frac{M_1}{0.7 M_{\odot}}\right)^{-0.85} \left(\frac{\dot{M}_{\rm acc}}{10^{18}~{\rm g/s}}\right)^{0.22} \\
+ 7.29 \times 10^{-4} \left(\frac{M_1}{0.7 M_{\odot}}\right)^{0.8} \left(\frac{\dot{M}_{\rm acc}}{10^{18}~{\rm g/s}}\right)^{1.0},
\end{split}
\label{height}
\end{equation}
where $R$ = $R_{1}$ \citep[see equation (2) in][]{pat85CVXrayemission2}. 
The accretion rate onto the WD, $\dot{M}_{\rm acc}$, is 
calculated with 
\begin{equation}
L_{\rm BB} = \frac{G M_1 \dot{M}_{\rm acc}}{2 R_{1}}~[{\rm ergs/s}],
\label{xray-lumi}
\end{equation}
where $G$ is the gravitational constant. 
Using equations (\ref{Rin-Rbb}), (\ref{height}), and 
(\ref{xray-lumi}), we derive the mass-radius relation 
of the WD to be 
\begin{equation}
\begin{split}
1.52 \times 10^{-6} \left(\frac{R_1}{10^8~{\rm cm}}\right)^{4.22} \left(\frac{M_1}{0.7 M_{\odot}}\right)^{-1.07} \left(\frac{L_{\rm BB}}{10^{34}~{\rm erg/s}}\right)^{-0.78}\\ 
+ 7.97 \times 10^{-8} \left(\frac{R_{1}}{10^8~{\rm cm}}\right)^{5} \left(\frac{M_1}{0.7 M_{\odot}}\right)^{-0.2}\\ 
= \left(\frac{L_{\rm BB}}{10^{34}~{\rm erg/s}}\right) \left(\frac{T_{\rm BB}}{10^{5}~{\rm K}}\right)^{-8}. 
\end{split}
\label{MR-relation}
\end{equation}
The observed quantities on BJD 2459551 are 
$L_{\rm BB}$ = (1.7$\pm$1.1)$\times$10$^{34}$~erg~s$^{-1}$ 
and $T_{\rm BB}$ = (3.5$\pm$0.1)$\times$10$^{5}$~K, where
the quoted errors are the combined ones for 
the 90\% confidence intervals of the distance to the source 
and the estimates from the SED modeling 
(Table \ref{parameter-211202}).
We applied these values to equation (\ref{MR-relation}) and 
obtained the possible range of $R_{1}$ for the given mass 
$M_1$, which is displayed in Figure \ref{wd-mass}. 
Combining the range with the theoretical mass-radius 
relation of the ONe WD with no rotation 
\citep[e.g.,][]{kas19magneticwind}, 
we conclude that the possible range of the WD mass 
in MASTER J0302 is 1.15--1.34~$M_{\odot}$ 
(Figure~\ref{wd-mass}).
The theoretical mass-radius relation  is known not 
to vary significantly as long as the spin period of 
the WD is longer than 10~s, which is usually the case 
for CVs \citep{lon03wzsgeFUSE,yua16gkper}. 
Hence, the uncertainty in applying the theoretical 
relation to our source should be limited. 

The estimated mass is $\sim$1.5-times heavier than 
the average WD mass in CVs, $\sim$0.8 $M_{\odot}$ 
\citep{lit08eclCV,sav11mass}.
We note that the actual WD mass may be even larger than 
this value. 
If we have overestimated the radiation efficiency of X-rays 
in equation (\ref{xray-lumi}), equation (\ref{MR-relation}) 
is modified and the possible range on the WD radius-mass 
plane (shaded area in Figure \ref{wd-mass}) would shift 
toward the lower radius. This leads to a higher mass 
in conjunction with the theoretical mass-radius relation, 
which is a decreasing function of the radius $R_{1}$.
The temperature $T_{\rm BL}$ of the optically thick 
boundary layer is approximated as 
\begin{equation}
T_{\rm BL} = 2.16 \times 10^5 \left(\frac{M_{1}}{0.7M_{\odot}}\right)^{0.86} 
\left(\frac{\dot{M}_{\rm acc}}{10^{18}}\right)^{0.18}~[{\rm K}], 
\label{BL-temp}
\end{equation}
\citep[see equation (3)][]{pat85CVXrayemission2}, 
and positively correlated with the WD mass.
The hypothesis that MASTER J0302 has a massive WD 
may naturally explain the observed high blackbody 
temperature.

Our analyses suggest that the WD in MASTER J0302 is 
an ONe WD with a mass very close to the Chandrasekhar mass.
If the WD mass is larger than 1.32$M_{\odot}$, 
the secondary star must be heavier than 0.08$M_{\odot}$ 
according to the assumed mass ratio (see Introduction). 
The mass transfer rate from the secondary star 
is $\sim$10$^{-10}$~$M_{\odot}$~yr$^{-1}$ \citep{kal16evolution}.
In this case, the WD in this source may become 
a neutron star through an accretion-induced collapse (AIC) 
event within $\sim$Gyr 
\citep[see][and references therein]{wan20aic}, providing 
that the ONe WD grows due to the accretion from 
the secondary star.
Then, after the AIC event, the system will likely 
become a binary system composed of a rapidly-rotating 
neutron star and a very light companion star, which is 
similar to a black-widow pulsar \citep{fru88blackwidow}.

The number of the discovered WZ Sge-type DNe is still 
limited. 
However, a significant observation bias is likely to be 
present because the typical outburst interval of 
WZ Sge-type DNe is $\sim$10 years \citep{kat15wzsge} 
and they would not be found during quiescence. 
Indeed, in the standard binary evolution scenario of CVs, 
WZ Sge-type DNe are the majority of the CV population 
\citep{kni11CVdonor,kal16evolution}.
The yet undiscovered population of WZ Sge-type DNe may 
contribute to the birth-rate of the binary system 
consisting of a millisecond pulsar and 
a low-mass star \citep{tau13aic}.
The observed space density of WZ Sge-type DNe is 
$\sim$10$^{-6}$~pc$^{-3}$ \citep{pal20SpaceDensity}. 
Currently, there are no WZ Sge-type systems except for 
MASTER J0302, in which we find observational evidence of 
a massive ONe WD.
The fraction of the objects like MASTER J0302 to 
WZ Sge-type DNe is at least $\sim$1/300. 
The event rate of AIC from MASTER J0302-like objects 
is roughly estimated as $\sim$10$^{-5}$~yr$^{-1}$~gal$^{-1}$.
This may account at least a fraction for 
the theoretically-predicted birth rate of 
millisecond pulsars \citep{hur10AICrate}.

\section{Conclusions}

\textcolor{black}{
MASTER J0302 exhibited an outburst in 2021--2022. 
This object was identified as a WZ Sge-type DN 
by detection of early and ordinary superhumps and 
double-peaked emission lines at the early stage of 
its outburst, though the outburst amplitude was 
$\sim$10 mag, extremely high (see section 3.1).
We analyzed the X-ray data taken with {\it NICER} 
and {\it NuSTAR}. 
The observed maximum X-ray luminosity was 
$\sim$10$^{35}$~erg~s$^{-1}$.
The {\it NICER} soft-X-ray observations enabled us 
to detect a blackbody component with a temperature of 
$\sim$30~eV and a luminosity of $\sim$10$^{34}$~erg~s$^{-1}$ 
around the outburst maximum, which rapidly declined. 
This was interpreted as the emission from the optically-thick 
boundary layer. 
Also, the bremsstrahlung emission with a temperature of 
$\sim$7~keV would originate from the optically-thin gas 
remaining even in outbursts. 
The partial absorption of soft X-rays by dense gas 
with $N_{\rm h}$ of $\sim$10$^{22}$~cm$^{-2}$ implies that 
outflows from the disk covered the X-ray emitting plasma 
(see sections 3.2 and 4.1).
These observational features were consistent with 
WZ Sge-type DNe; however, prominent oxygen and neon 
emission lines were extraordinary.
It is unlikely that the secondary star provided 
oxygen and neon-rich gas. 
The enrichment could be associated with past nova eruptions 
if the central WD is an ONe WD, though this is still in debate 
(section 4.2). 
We estimated the WD mass from the luminosity and temperature 
of a blackbody component originating from the boundary layer 
to be as heavy as 1.15--1.34$M_{\odot}$.
The system might evolve to a binary including a millisecond 
pulsar via AIC within another $\sim$Gyr (section 4.3). 
}

\section*{acknowledgments}

M.~Kimura acknowledges support by the Special Postdoctoral 
ResearchersProgram at RIKEN.  
We are thankful to Dr.~Masaaki Sakano for his English 
proofreading.
We are grateful to the many amateur observers in VSNET 
for providing the optical light curve presented in 
this paper.
We acknowledge Dr.~Keisuke Isogai who took optical spectra 
presented in this paper.
This work was financially supported by Japan Society for 
the Promotion of Science Grants-in-Aid 
for Scientific Research (KAKENHI) Grant Numbers
JP20K22374 (MK), JP21K13970 (MK), JP20K04010 (KK), 
JP20H01904 (KK), JP22H00130 (KK), JP22K03688 (TS), 
JP22K03671 (TS), JP20H05639 (TS), JP21J22351 (YT). 
We thank the anonymous referee for 
helpful comments.



\newcommand{\noop}[1]{}

\appendix

\section{Supplementary tables and figures}

\restartappendixnumbering

\begin{figure*}[htb]
\begin{minipage}{0.49\hsize}
\plotone{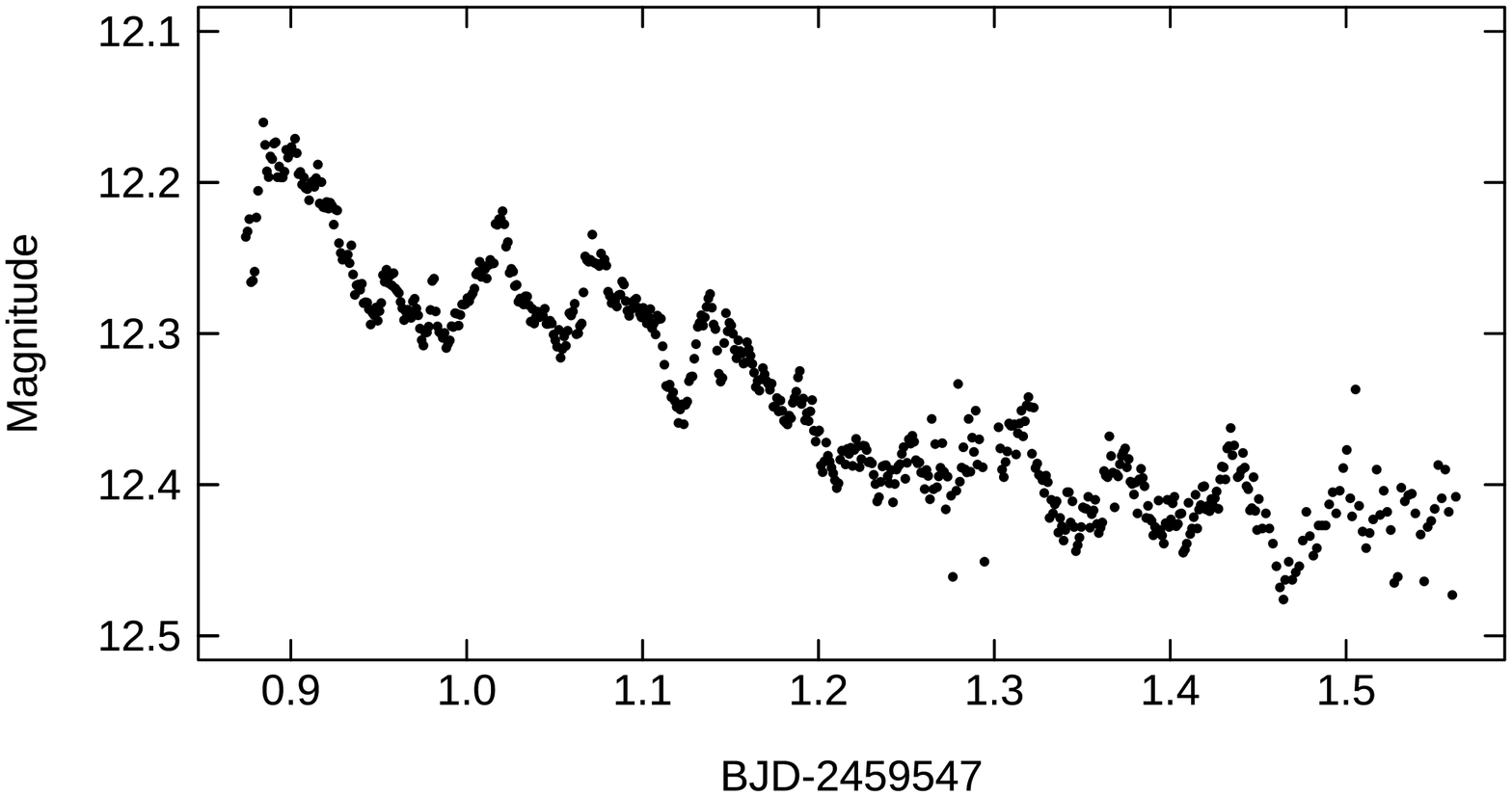}
\end{minipage}
\begin{minipage}{0.49\hsize}
\plotone{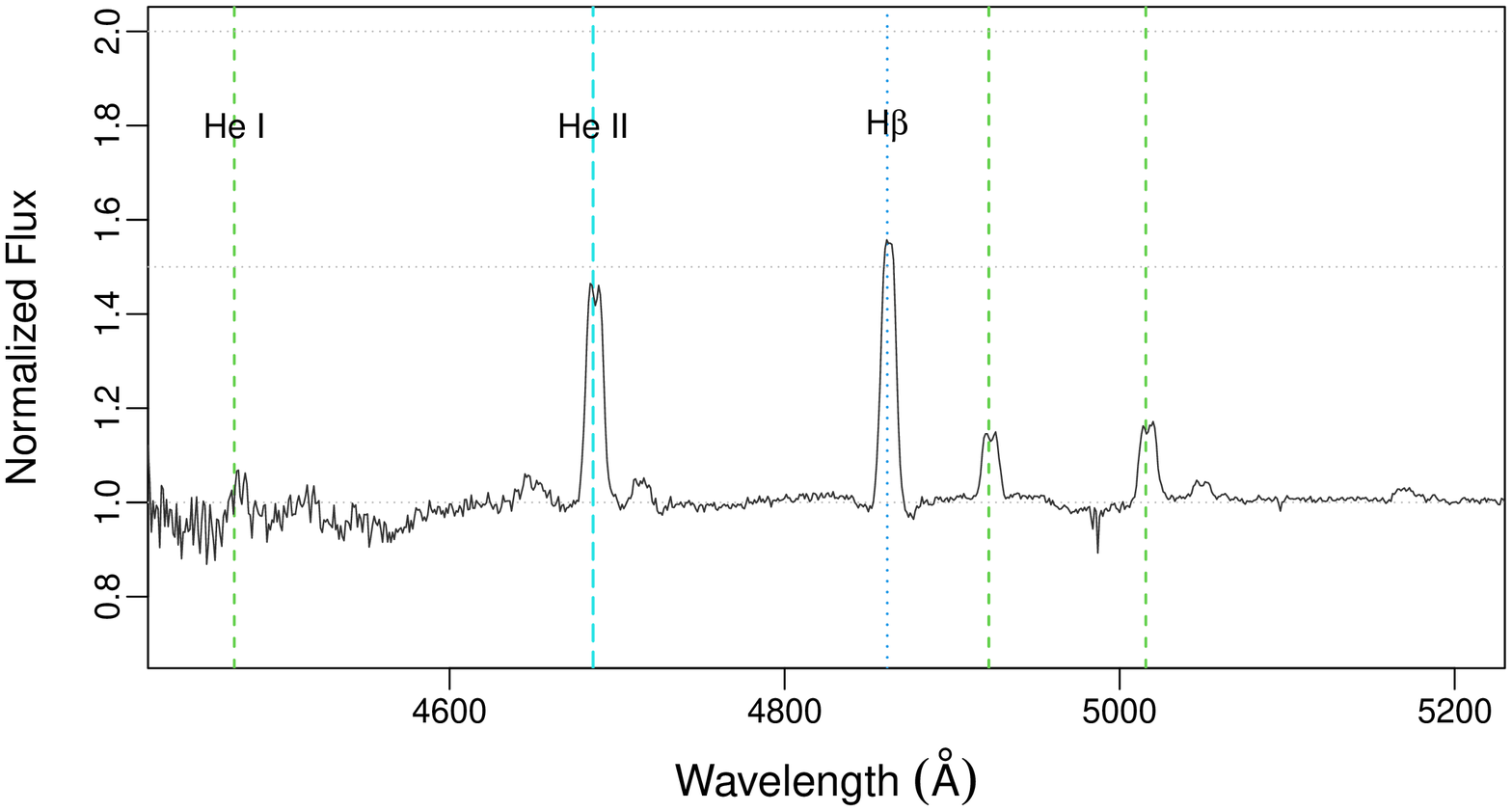}
\end{minipage}
\caption{
(Left) An example of light curves of early superhumps during the 2021--2022 outburst of MASTER J0302, whose period is identical with the orbital period (Tampo et al.~in prep.).
(Right) An example of optical spectra in which double-peaked emission lines are dominant on BJD 2459547 \citep{iso21afpi}. This was taken by Kyoto Okayama Optical Low-dispersion Spectrograph with an Integral Field Unit (KOOLS-IFU; \cite{kools}) mounted on the 3.8-m telescope Seimei at Okayama Observatory, Kyoto University \citep{Seimei}. The blue dot, green dashed, and cyan dashed lines represent the central wavelength of Balmer, He I, and He II lines, respectively.
}
\label{earlySH-spectrum}
\end{figure*}

\begin{table*}[htb]
\caption{Log of observations of MASTER J0302 with {\it NICER}.  }
\label{nicer-log}
\begin{center}
\begin{tabular}{cccccc}
\hline
NICER~ObsID & ${\rm Start}^{*}$ & ${\rm End}^{*}$ & On-source time$^{\dagger}$ & Average rate$^{\ddagger}$ \\ \hline
4202450101 & 547.6613 & 548.4464 & 8181 & 13.8 \\
4202450102 & 548.6296 & 549.4791 & 5199 & 5.9 \\
4202450103 & 549.5335 & 550.4653 & 9169 & 3.9 \\
4202450104 & 550.5297 & 551.4982 & 9432 & 1.9 \\
4202450105 & 551.5373 & 552.4495 & 4538 & 1.6 \\
4202450106 & 552.5053 & 553.4805 & 10101 & 1.4 \\
4202450107 & 553.5360 & 554.4495 & 8154 & 1.1 \\
4202450108 & 554.5045 & 555.4348 & 9116 & 1.0 \\
4202450109 & 555.7957 & 556.4547 & 3716 & 0.9 \\
4202450110 & 557.1680 & 557.3630 & 2446 & 0.7 \\
4202450111 & 557.6674 & 558.3249 & 3938 & 0.7 \\
4202450112 & 558.5711 & 559.2946 & 5933 & 0.7 \\
4202450113 & 559.8632 & 559.8779 & 1267 & 0.7 \\
4202450114 & 561.0258 & 561.1776 & 5714 & 0.7 \\
4202450117 & 571.6462 & 571.6506 & 375 & 0.5 \\
4202450118 & 572.9937 & 573.0676 & 1754 & 0.5 \\
4202450119 & 573.5130 & 573.5213 & 712 & 0.5 \\
4202450120 & 574.9912 & 575.0070 & 1369 & 0.6 \\
4202450121 & 576.7971 & 576.9312 & 2247 & 1.0 \\
4202450122 & 578.0803 & 578.0951 & 1211 & 0.8 \\
4202450123 & 578.9859 & 579.0042 & 1571 & 0.7 \\
4202450124 & 579.6328 & 579.6470 & 667 & 0.4 \\
4202450125 & 600.2890 & 600.2986 & 833 & 0.7 \\
4202450126 & 600.6161 & 600.6252 & 789 & 0.5 \\
4202450127 & 601.5152 & 601.9802 & 1842 & 0.9 \\
4202450128 & 603.2009 & 603.2064 & 470 & 0.6 \\
4202450129 & 604.6126 & 605.2021 & 1681 & 1.8 \\
4202450130 & 605.5150 & 606.4878 & 3069 & 1.6 \\
4202450131 & 606.5463 & 606.7459 & 1693 & 1.5 \\
4202450132 & 607.8387 & 607.8476 & 774 & 0.9 \\
4202450133 & 608.5486 & 609.3911 & 2338 & 1.4 \\
4202450134 & 609.5164 & 610.1706 & 1109 & 0.8 \\
4202450135 & 610.8065 & 610.8207 & 1219 & 1.0 \\
4202450136 & 611.5806 & 611.5934 & 1094 & 0.8 \\
4202450137 & 613.1933 & 613.2737 & 2706 & 1.0 \\
4202450138 & 614.1603 & 614.1702 & 851 & 1.8 \\
4202450139 & 616.8054 & 617.1925 & 260 & 0.9 \\
\hline
\multicolumn{5}{l}{$^{*}$BJD$-$2459000.0.}\\
\multicolumn{5}{l}{$^{\dagger}$Units of seconds.}\\
\multicolumn{5}{l}{$^{\ddagger}$NICER count rate in 0.3--7~keV in units of counts/sec.}\\
\end{tabular}
\end{center}
\end{table*}

\begin{table*}[htb]
\caption{Log of observations of MASTER J0302 with {\it NuSTAR}.  }
\label{nustar-log}
\begin{center}
\begin{tabular}{cccccc}
\hline
NuSTAR~ObsID & ${\rm Start}^{*}$ & ${\rm End}^{*}$ & On-source time$^{\dagger}$ & Average rate$^{\ddagger}$ \\ \hline
90701341002 & 550.4917 & 551.2005 & 36218 & 0.4 \\
\hline
\multicolumn{5}{l}{$^{*}$BJD$-$2459000.0.}\\
\multicolumn{5}{l}{$^{\dagger}$Units of seconds.}\\
\multicolumn{5}{l}{$^{\ddagger}$NuSTAR count rate in 3--79~keV in units of counts/sec.}\\
\end{tabular}
\end{center}
\end{table*}

\begin{figure*}[htb]
\begin{minipage}{0.24\hsize}
\plotone{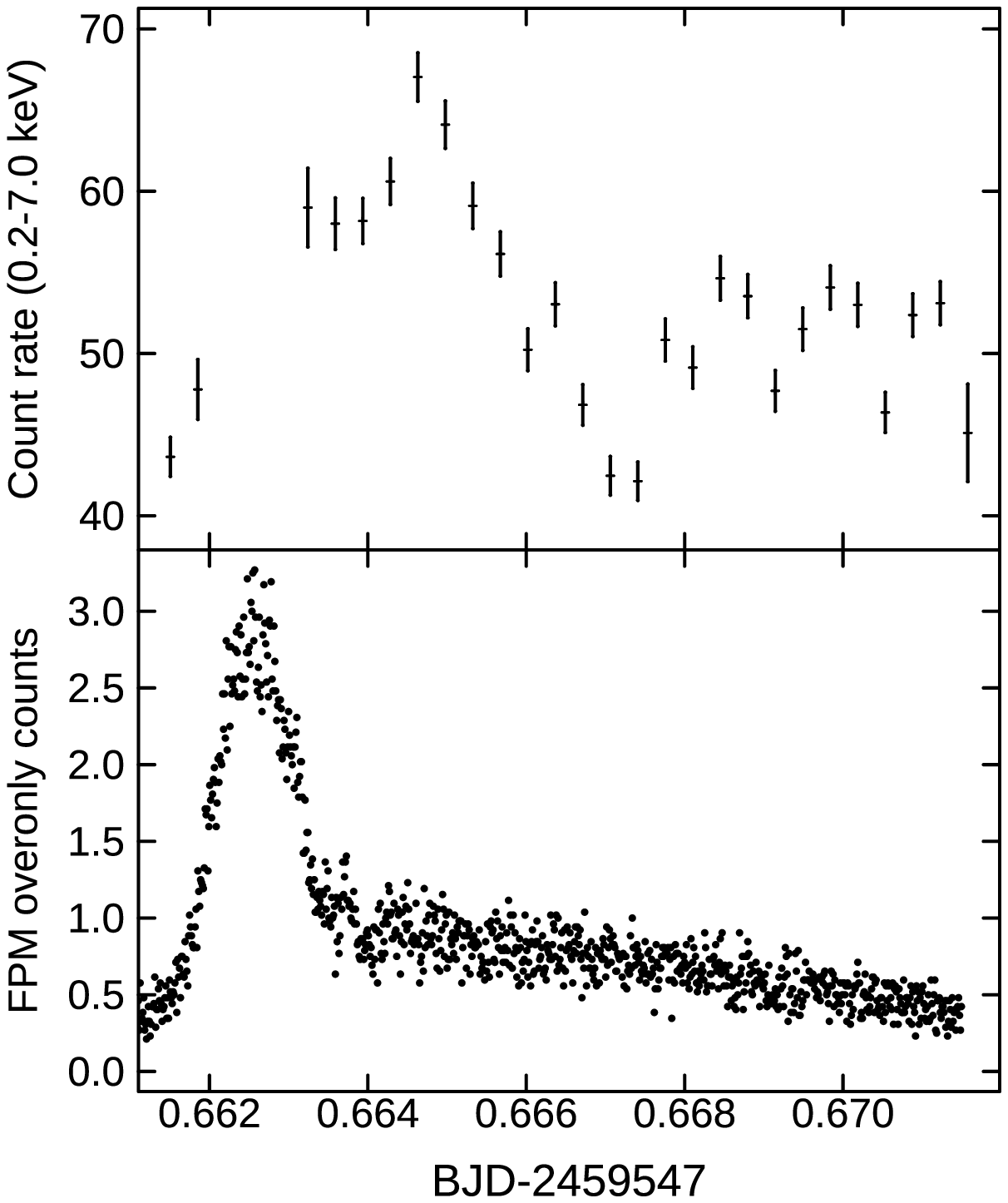}
\end{minipage}
\begin{minipage}{0.24\hsize}
\plotone{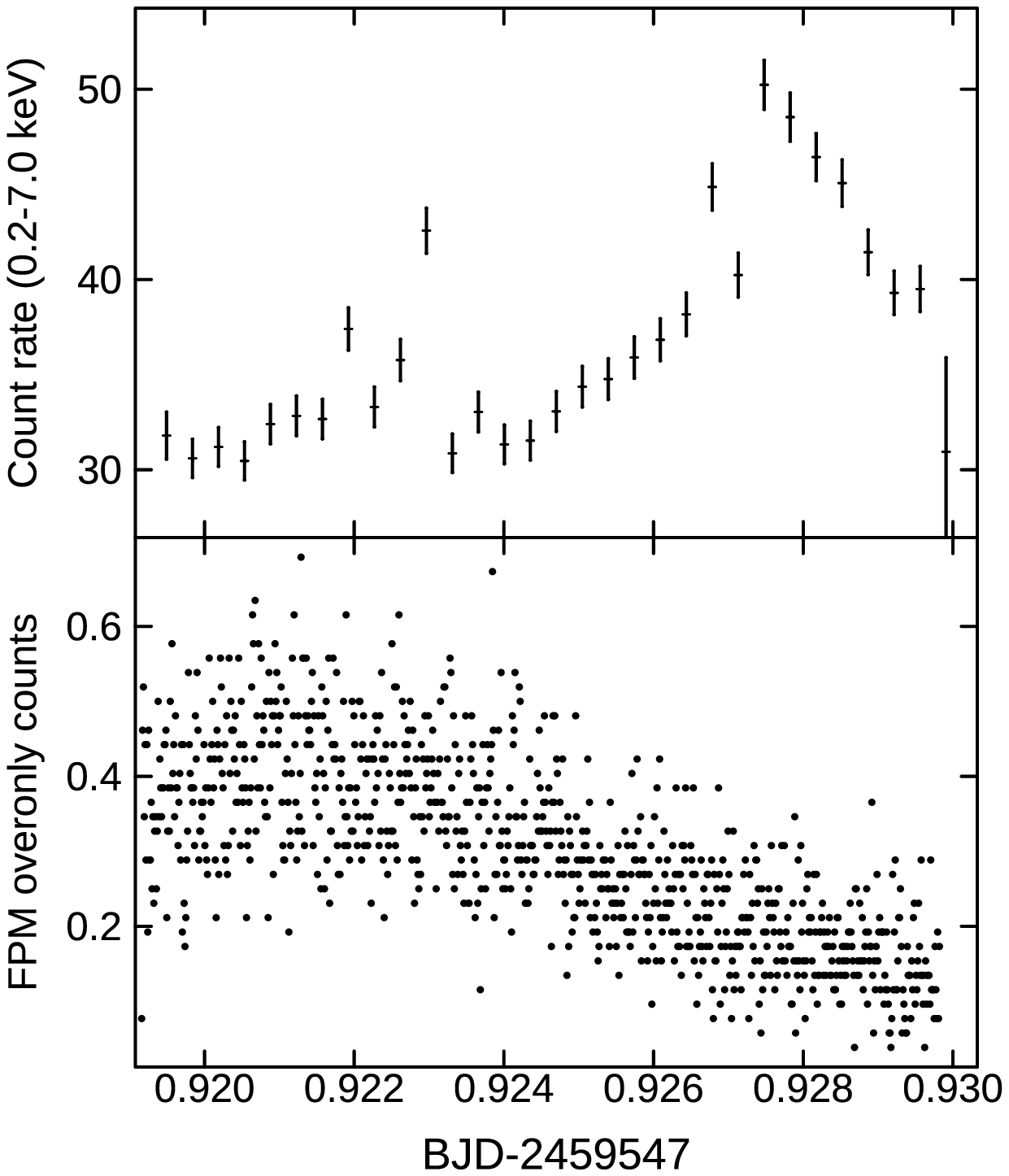}
\end{minipage}
\begin{minipage}{0.24\hsize}
\plotone{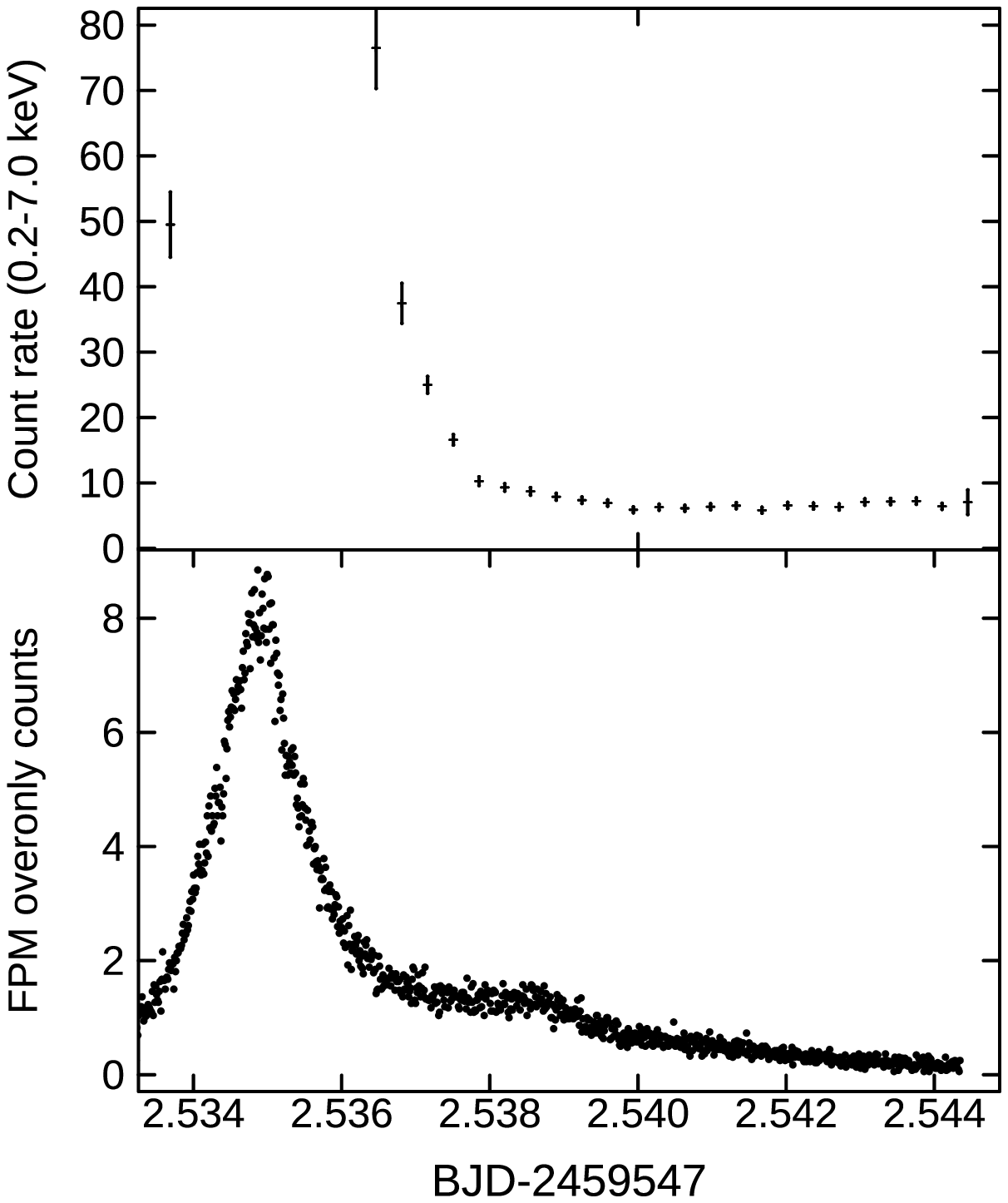}
\end{minipage}
\begin{minipage}{0.24\hsize}
\plotone{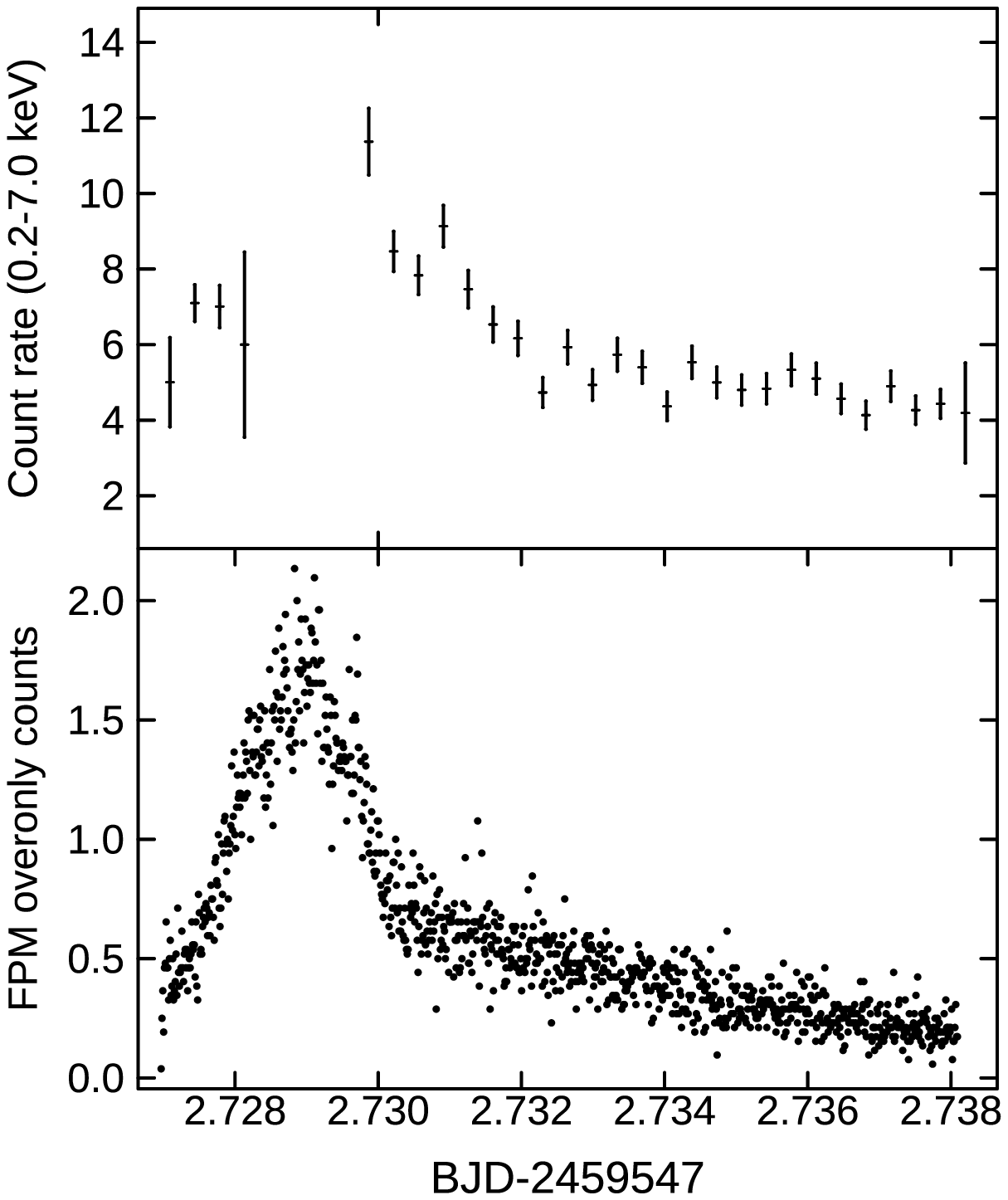}
\end{minipage}
\caption{
Examples of X-ray variability in the 2021--2022 outburst of MASTER J0302. The upper and lower panels of each figure denote the {\it NICER} count rate in the 0.2--7.0 keV and FPM overonly count which represents the average per-detector overshoot rate, respectively. 
Most of flares seemed to be caused by charged particles. When high energy charged particles such as cosmic rays pass through the {\it NICER} SDDs, many charges deposited in SDDs along its path are registered as an overshoot event.
The flare in the second figure from the left seemed to be intrinsic to MASTER J0302 because there is no flare in FPM overonly count, and was dominant in the softer energy range.
}
\label{variability}
\end{figure*}

\begin{figure*}[htb]
\begin{minipage}{0.49\hsize}
\plotone{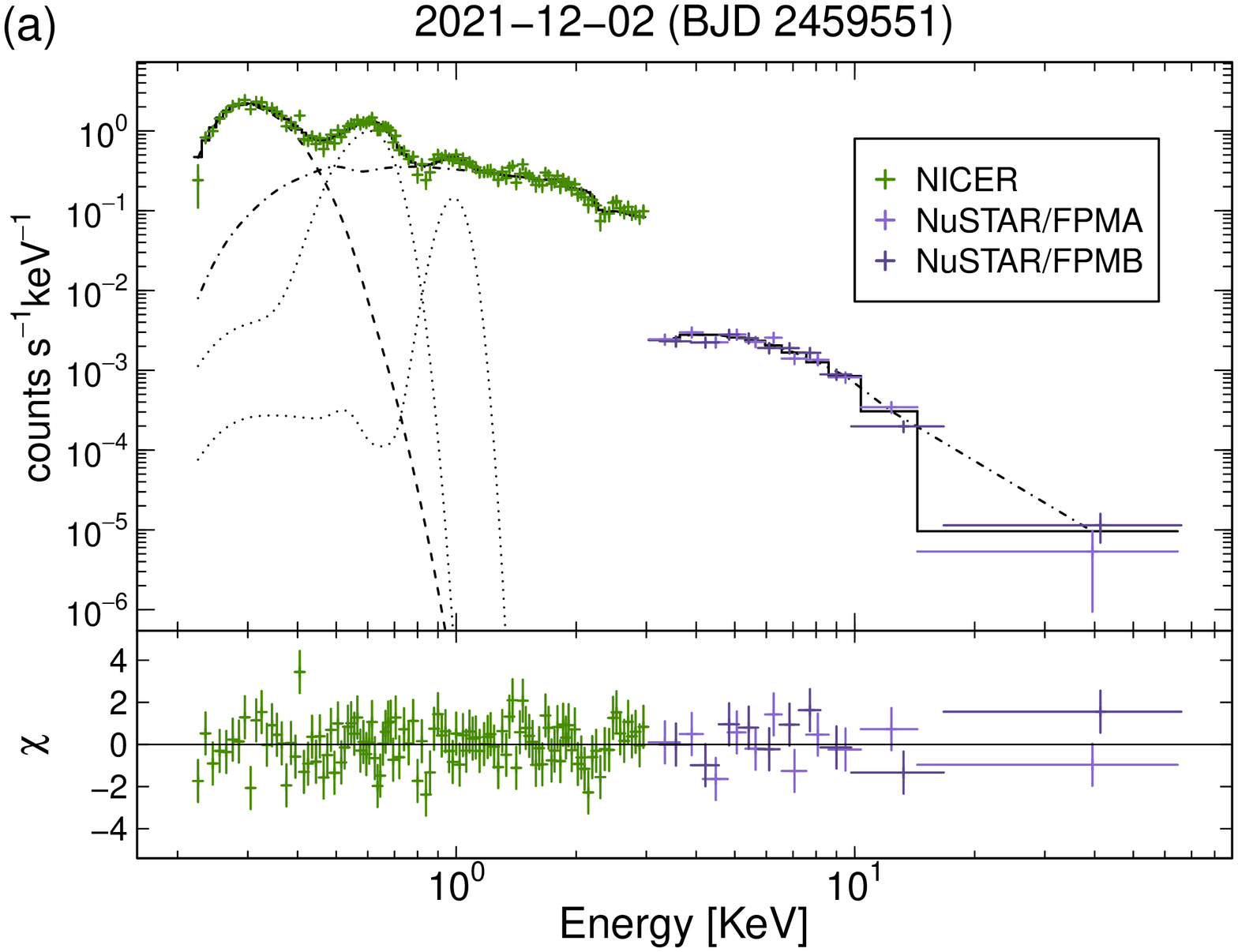}
\end{minipage}
\begin{minipage}{0.49\hsize}
\plotone{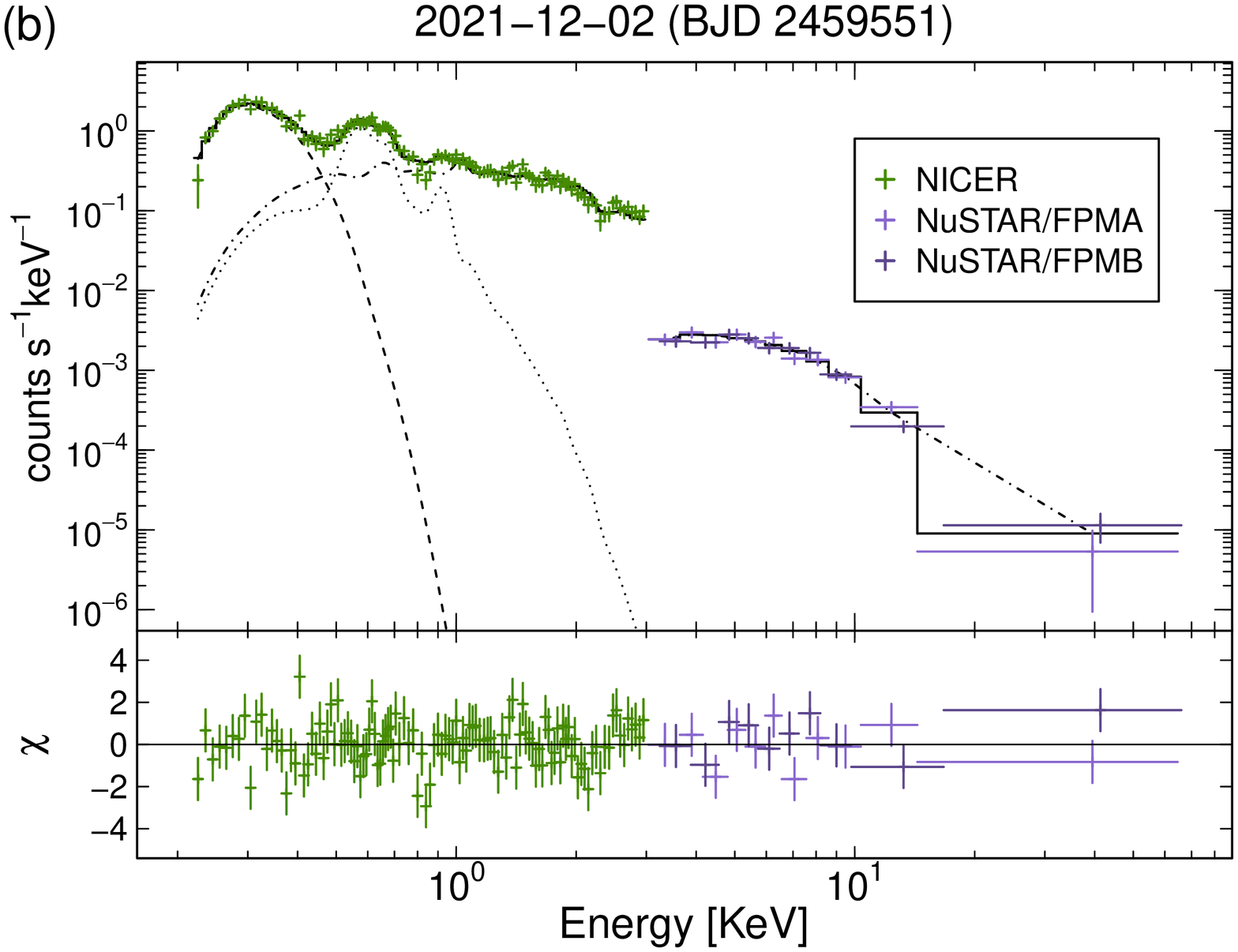}
\end{minipage}
\caption{
Broad-band X-ray data of MASTER J0302 on BJD 2459551 (2021 December 2) during the outburst. The green crosses represent the {\it NICER} data. The purple and dark purple crosses represent the {\it NuSTAR} FPMA and FPMB data, respectively. 
(a) The data with the best-fit spectral model of \texttt{Tbabs*pcfabs*(bbody+gaussian+gaussian+bremsstrahlung)}. The dashed line and dot-dashed line represent the model emission from a blackbody source and that from bremsstrahlung source, respectively. The two dot lines denote the model emission from oxygen and neon lines. The solid line is a sum of the model emission. 
(b) The data with the best-fit spectral model of \texttt{Tbabs*pcfabs*(bbody+vapec+vapec)}. The dashed line, dot line, and dot-dashed line represent the model emission from a blackbody source, a low-temperature and a high-temperature collisionally-ionized plasma, respectively. The solid line is a sum of the model emission.
}
\label{spec211202-ld}
\end{figure*}



\end{document}